\documentclass[twocolumn]{aastex631}
\usepackage{color}

\shorttitle{CSST $\times$ LSST}
\shortauthors{Cao et al.}

\begin{document}
\title{Forecast of joint analysis of cosmic shear and supernovae magnification from CSST and LSST}

\newcommand{\BNUifaa}{\affiliation{Institute for Frontier in Astronomy and Astrophysics, Beijing Normal University, Beijing, 102206, People's Republic of China}}
\newcommand{\bnu}{\affiliation{Department of Astronomy, Beijing Normal University, Beijing, 100875, People's Republic of China}}
\newcommand{\shao}{\affiliation{Shanghai Astronomical Observatory (SHAO), Nandan Road 80, Shanghai, 200030, People's Republic of China}}
\newcommand{\naoc}{\affiliation{Key Laboratory of Space Astronomy and Technology, National Astronomical Observatories, Chinese Academy of Sciences, Beijing, 100101, People's Republic of China}}
\newcommand{\pku}{\affiliation{Kavli Institute for Astronomy and Astrophysics, Peking University, Beijing, 100871, People's Republic of China}}

\author{Ye Cao}
\BNUifaa
\bnu

\author{Bin Hu}
\correspondingauthor{E-mail: bhu@bnu.edu.cn}
\BNUifaa
\bnu

\author{Ji Yao}
\shao

\author{Hu Zhan}
\naoc
\pku

\begin{abstract}
Cosmic shear and cosmic magnification reflect the same gravitational lensing field. Each of these two probes are affected by different systematics. 
We study the auto- and cross-correlations of the cosmic shear from the China Space Survey Telescope (CSST) and cosmic magnification of supernovae from Large Synoptic Survey Telescope (LSST). We want to answer, to what extent, by adding the magnification data we can remove the systematic bias in cosmic shear measurement. 
We generate the mock shear/magnification maps based on the correlation between of different tomographic bins. After obtaining the corrected power spectra, we adopt the Markov Chain Monte Carlo (MCMC) technique to fit the theoretical models, and investigate the constraints on the cosmological and nuisance parameters. 
We find that the with only cosmic shear data, there are $1\sigma$ bias in $\sigma_8$ and intrinsic alignment model parameters. By adding the magnification data, we are able to remove these biases perfectly. 
\end{abstract}


\section{Introduction}
Weak gravitational lensing is a powerful observational tool for investigating cosmological and astronomical questions such as gravity, dark matter, dark energy, and the cosmic large-scale structure \citep{Kaiser92,Refregier03,Mandelbaum18,Blake20}. The weak gravitational lensing effect can distort the shapes of galaxy images, it is commonly referred to as cosmic shear \citep{Albrecht06,Peacock06}. The cosmic shear can be used to estimate the spatial distribution of gravitating matter along the line of sight \citep{Kilbinger15}, and is sensitive to the amplitude and shape of the matter power spectrum \citep{Van01,Kilbinger15,Asgari21}. Therefore, the cosmic shear is commonly used to detect weak gravitational lensing. Numerous ongoing and planned telescopes with cosmic shear measurement are well known, e.g. Euclid space telescope \citep{Laureijs11} and China Space Station Telescope (CSST) \citep{Zhan11,Zhan21}. These powerful surveys are expected to make great improvements on cosmological and astronomical objectives. However, the observed cosmic shear signal is contaminated by the intrinsic alignments and systematics. Therefore, we must build related models and parameters to eliminate their effects, which makes the constraint results looser on the cosmological parameters.

Type Ia supernovae (SNIa) are widely studied as powerful probes on the expansion history of the universe \citep{Knop03,Barris04,Riess04,Huterer05}. by using the luminosity distance and redshift duality relationship. 
Because the luminosity distance of SNIa is also affected by the lensing field, once the number density of SNIa reaches a threshold, eg. $38$ per squared degree, one can also use the magnification information to study lensing field on degree scales \citep{Jain02,Cooray06}. 
The reliable detection of cosmic magnification was performed by the Sloan Digital Sky Survey (SDSS)\citep{Scranton05}. In recent years, a number of SNIa surveys are being planned or performed, such as the NASA/DOE Joint Dark Energy Mission (JDEM\footnote{\url{http://jdem.gsfc.nasa.gov/}}) and the Large Synoptic Survey Telescope (LSST)\citep{LSST09,Ivezic19}.

The CSST is a 2-meter multi-band space telescope, which will share the same orbit as the China Manned Space Station. The large-scale multicolor imaging and slitless spectroscopic observations are the two major projects. The CSST has excellent performance such as a large field of view (approximately 1.1 $\rm deg^2$), high spatial resolution (around 0.15 arcsec in the 80\% energy concentration region), sensitivity to faint magnitudes, and wavelength coverage from NUV to NIR \citep{Zhan21,Gong19,Cao22}. The CSST will cover more then 17,500 $\rm deg^2$ sky area in 10 years, and explore some important cosmological and astronomical scientific objectives. The LSST is a ground-based wide-field imaging telescope with an effective aperture of 6-8 meter. the LSST will obtain multi-band images over half the sky in 10 years, and will detect millions of supernova \citep{Olivier07,LSST09,Ivezic19}. The unprecedented number of supernova allows us to detect and calibrate the systematic effect \citep{Wood-Vasey06,Olivier07}. Therefore, the CSST and LSST are expected to be powerful surveys for investigating the fundamental cosmological and astronomical questions such as gravity, dark matter, and dark energy, the cosmic large-scale structure, etc.

In this work, we enhance the cosmological parameter constraints of CSST shear survey through a joint analysis with LSST magnification maps. We generate the auto- and cross-power spectra of the magnification-magnification, magnification-shear and shear-shear from LSST and CSST surveys. We analyze LSST Wide-Fast-Deep Survey region (WFD) areas covering about 18,000 $\mathrm{deg^2}$ and CSST main sky survey areas covering about 18,000 $\mathrm{deg^2}$, and their overlap areas greater than 7,000 $\mathrm{deg^2}$. In order to reduce degeneracies of redshift dependent parameters and improve the constraints on cosmological parameters, we divide the example sources into several photometric redshifts (photo-$z$) tomographic bins. We generate the realizations of correlated magnification maps and shear maps using the theoretical power spectra by Healpy package \citep{Gorski05,Zonca19}. After correcting power spectrum with the mode coupling matrix and removing noise, we obtain the magnification-magnification and magnification-shear power spectra from $30\le \ell \le 383$ and the shear-shear power spectra from $30\le \ell \le3,000$ (a conservative scale cut with $30\le \ell \le1,000$). Besides, we analyze the covariance matrix of the power spectra, and perform a Markov Chain Monte Carlo (MCMC) analysis to constrain the cosmological parameters and intrinsic alignment model.

This paper is organized as follows: in Section \S\ref{SNe}, we describe the supernovae magnification measurement and their theoretical models, and generate the mock data for LSST. In Section \S\ref{shear}, we introduce the "TATT" (tidal alignment and tidal torquing) model for the cosmic shear measurement, describe the correlation between supernovae magnification and cosmic shear, and create the mock CSST mock shear data. In Section \S\ref{spectrum}, we describe the final power spectrum measurements and corrections. In Section \S\ref{spec_result}, we present the measured power spectra and use a theoretical model to fit them by MontePython code \citep{Audren13}. We finally summarize the results in Section \S\ref{conclusions}. Throughout this paper, we assume a fiducial flat $\Lambda \mathrm{CDM}$ cosmological model with $\Omega_{\mathrm{m}}$ = 0.3, $\Omega_{\mathrm{b}} h^2$ = 0.0245, $\sigma_{8}$ = 0.8, $n_s$ = 0.95 and $h$ = 0.7.

\section{Supernovae magnification}
\label{SNe}

The main survey strategy of LSST is called WFD with limiting magnitude of $r \sim 24.16$). It covers about 18,000 $\mathrm{deg^2}$ and uses about 90\% of the observing time \citep{LSST17}. Hundreds of thousands of SNIa will be discovered in the WFD field over a 10-year survey. Hence it is expected to obtain supernovae magnification measurements for the weak lensing study. In this section, we discuss the magnification power spectra and mock images for the LSST.

\subsection{Model of Supernovae Magnification}
\label{subsec:SNe_model}

The luminosity distance of a given SNIa at a redshift $z$ and located in the direction $\hat{\mathbf{n}}$, $D_{\mathrm{L}}(z, \hat{\mathbf{n}})$, is magnified by weak lensing of all mater from us to the SNIa, so the lensed luminosity distance $D_{\mathrm{L}}^{\mathrm{len}}(z, \hat{\mathbf{n}})$ is related to the ture luminosity distance $D_{\mathrm{L}}(z,\hat{\mathbf{n}})$ by \citep{Shapiro10}:
\begin{equation}
D_{\mathrm{L}}^{\mathrm{len}}\left(z, \hat{\mathbf{n}}\right) = \frac{D_{\mathrm{L}}\left(z, \hat{\mathbf{n}}\right)}{\sqrt{\mu \left(z, \hat{\mathbf{n}}\right)}},
\label{eq:D_L}
\end{equation}
where $\mu \left(z, \hat{\mathbf{n}}\right)$ is the lensing magnification at redshift $z$ and in the direction $\hat{\mathbf{n}}$. Since $\left\langle \mu \left(z\right) \right\rangle=1$ at redshift $z$ for the all sky, the average observed luminosity distance $\overline D_{\mathrm{L}}(z)$ of over large samples at redshift $z$ is equal to the true luminosity distance at this redshift. The magnification fluctuations is given by
\begin{equation}
\label{eq:mag_flu}
\delta_{\mu} = \left[\frac{\overline D_{\mathrm{L}}(z)}{D_{\mathrm{L}}^{\mathrm{len}}\left(z, \hat{\mathbf{n}}\right)}\right]^2 -1 .
\end{equation}

In the limit of weak lensing ($\mu , \kappa \ll 1$), the lensing magnification is directly related to the lensing convergence by $\mu \simeq 1+2\kappa$. Then we discuss the detailed estimates of the theoretical angular spectrum using the gravitational lensing potential $\phi$. We can relate the lensing convergence to the lensing convergence by
\begin{equation}
\kappa=\frac{1}{2} \ell(\ell+1) \phi.
\end{equation}
We then define the lensing potential \citep{Hu01,Lewis06},
\begin{equation}
\phi(\hat{\mathbf{n}}) = -2 \int_{0}^{\chi_{\mathrm{H}}}d\chi_{\mathrm{H}} \int_{\chi}^{\chi_{\mathrm{H}}}d\chi' p\left(\chi' \right)\frac{f_{\Omega_K}(\chi'-\chi)}{f_{\Omega_K}(\chi')f_{\Omega_K}(\chi)} \Psi\left(\chi \hat{\mathbf{n}} ; \chi\right),
\end{equation}
where $\chi_{\mathrm{H}}$ denotes the distance to the horizon, $p\left(\chi \right)$ is the normalized distribution of SNIa, and $\int p\left(\chi \right)d\chi =1$. $\Psi\left(\chi \hat{\mathbf{n}} ; z\right)$ is the three-dimensional gravitational potential, and $f_{\Omega_K}(\chi) = \chi$ denotes the comoving angular diameter distance in flat universe. The comoving distance along the sight $\chi(z)$ is defined by
\begin{equation}
\chi(z) = D_{\mathrm{H}}\int_{0}^{z} \frac{H_0}{H(z')}dz',
\end{equation}
where $H_0 = 100h\ \rm{km\ s^{-1}\ Mpc^{-1}}$ denotes the Hubble parameter today, and $D_{\mathrm{H}}$ denotes the Hubble distance today, where $D_{\mathrm{H}} = c / H_0$. 

The lensing magnification can be decomposed into spherical harmonic coefficients by
\begin{equation}
\mu_{\ell m} =\int d \hat{\mathbf{n}} \mu(\hat{\mathbf{n}}) Y_{\ell m}^*(\hat{\mathbf{n}}).
\end{equation}
Assuming statistical isotropy, the angular power spectrum of magnification fluctuations can be written as
\begin{equation}
\left\langle \mu_{\ell m} \mu_{\ell^{\prime} m^{\prime}}^{*}\right\rangle =\delta_{\ell \ell^{\prime}} \delta_{m m^{\prime}} C_{\ell}^{\mu \mu}.
\end{equation}

We generally divide samples based on photo-$z$ distribution into different tomographic bins, which helps improve the constraint capabilities on cosmological parameters and reduces degeneracies of redshift dependent parameters \citep{Hu99}. For a flat universe, under the Limber-approximate \citep{Limber53}, the cross-power spectrum of signal between the $i$-th and $j$-th tomographic redshift bins is written as 
\begin{equation}
{C}^{ij}_{\rm \mu \mu}\left(\ell\right) = \int_{0}^{z_{\mathrm{H}}}dz\frac{d\chi}{dz}\frac{W_{\mu}^{i}(z)W_{\mu}^{j}(z)}{\chi^2}P_{\delta}(k=\ell/\chi,z),
\label{eq:cluu}
\end{equation}
Where, ${C}^{ij}_{\rm \mu \mu}\left(\ell\right)$ is the auto-power spectrum if $i=j$. $P_{\delta}(k,z)$ is the nonlinear matter power spectrum at redshift $z$, and we calculate it using the nonlinear Halofit model by the Cosmic Linear Anisotropy Solving System (CLASS) \citep{Blas11}. $W_{\mu}^{i}(z)$ is the lensing weighting function at redshift $z$ in the $i$-th tomographic redshift bin, and is given by
\begin{equation}
W_{\mu}^i(z) =\frac{3 \Omega_m H_0^2 \chi}{c^2 a} \int_{z}^{z_{\mathrm{H}}}d\chi' \frac{d\chi'}{dz}p^i\left(z' \right)\frac{\chi'-\chi}{\chi'},
\end{equation}
where $c$ is the speed of light, and $a=1/\left(1+z\right)$ is the scale factor.
The measured magnification power spectrum at a given multipole $\ell$ can be written by
\begin{equation}
 \hat{C}^{ij}_{\rm \mu \mu}\left(\ell\right)={C}^{ij}_{\rm \mu \mu}\left(\ell\right)+\delta_{i j}N^i_{\mathrm{\mu \mu}}(\ell) \;,   
\end{equation}
where $N^i_{\mathrm{\mu \mu}}$ is the noise spectrum when $i=j$, it depends on the uncertainty of the luminosity distance measurement, the number density of SNe, the width of the redshift bin, and so on.

\subsection{Mock SNIa Catalog}
\label{subsec:SNe_catalog}

There are two steps to obtain the simulated sky map, one is to construct a mock SNIa catalog, and the other is to generate simulated magnification images with that catalog. The mock SNIa catalog contain the information about the fiducial redshift, photo-$z$, fiducial luminosity distance, observed luminosity distance and positions of SNIa. After obtaining the mock SNIa catalog, we divide all the SNIa into five tomographic redshift bins and create a magnification fluctuations map based on the mock data by Eq. \ref{eq:mag_flu} for each tomographic redshift bin.

\begin{figure*}
\gridline{\fig{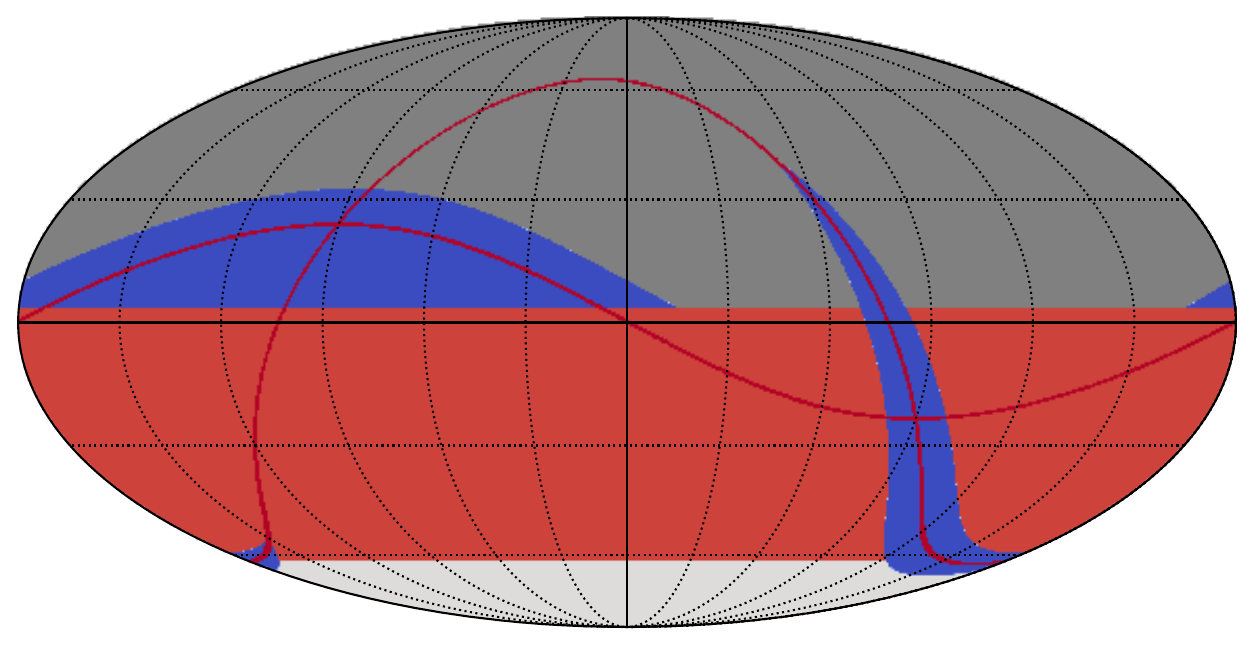}{0.49\textwidth}{}
          \fig{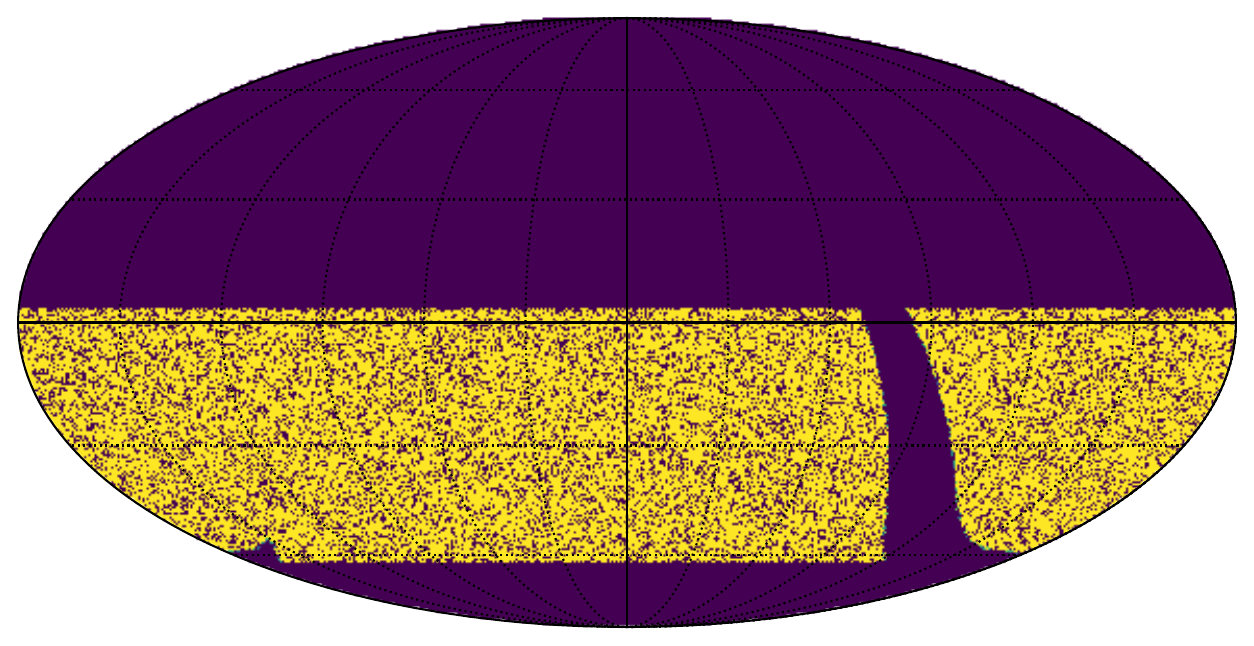}{0.49\textwidth}{}}
\caption{ $\it left$: The distribution of the SNIa on the sky for one simulated realization of the main survey. The sky is shown in HEALPix projection in equatorial over a 10-year time span. The red area is the main survey called “Wide-Fast-Deep (WFD)”, and the two blue regions with smaller number of SNIa are the North Ecliptic Spur (NES) region (on the left) and the Galactic Plane (on the right). The red lines show the Ecliptic and the Galactic equator. The regions around the South Celestial Pole (SCP) and the unobserved area are shown in white and gray, respectively. $\it right$: The masked image that removes the areas that do not contain measured data for the tomographic redshift bin with $0.65<z_{\mathrm{photo}}\le1.2$. The mask over the whole WFD field removes $\sim 8\%$ of the pixels.}
\label{fig:map0} 
\end{figure*}

LSST's observation strategy is to scan the sky deeply, widely and fast to produce data sets that simultaneously meet the majority of the science requirement. We show the four LSST survey fields in the left sub-panel of Fig. \ref{fig:map0}, and the sky is shown in HEALPix\footnote{\tt http://healpix.sf.net}\citep{Gorski05,Zonca19} projection in equatorial over a 10-year time span. The red area is the WFD, which is expected to contain $\sim 85\%$ of the total operation time. We construct mock catalog and generate magnification images for this region. The two blue regions with smaller number of SNIa are the North Ecliptic Spur (NES) region (on the left) and the Galactic Plane \citep{LSST17}. The regions around South Celestial Pole (SCP) and the unobserved area are shown in white and gray, respectively.

To rigorously test analysis pipelines, the LSST Dark Energy Science Collaboration (DESC) did a study called Data Challenge 2 (DC2), and generated a large and comprehensive set of image simulations \citep{LSST09,Sanchez20,Sanchez22}. \cite{Sanchez22} select a 15 $\mathrm {deg}^2$ area from the DC2 WFD and find 504 events. We create a simulation ($\sim 610, 000$ SNIa for the whole WFD field) that has the same SNIa number density as DC2 data.

In order to extract more information from the weak lensing data, we divide the redshift range into different tomographic redshift bins, and study the magnification power spectra of these tomographic redshift bins. 
For LSST survey, most of the redshift information of the SNe will come from photo-$z$ measurements of the SNe or its host galaxy.
Assuming that the true redshift distribution at a given photo-$z$, $z_{\rm{ph}}$, is given by the conditional probability distribution $p_{\rm t}(z \mid z_{\rm{ph}})$. For simplicity, we assume $p_{\rm t}(z \mid z_{\rm{ph}})$ is a Gaussian distribution function at each $z_{\rm{ph}}$ \citep{Ma06,Yao23,Miao23}, and can be expressed as 
\begin{equation}
p_{\rm t}\left(z \mid z_{\mathrm{ph}}\right)=\frac{1}{\sqrt{2 \pi} \sigma_z} \exp \left[-\frac{\left(z-z_{\mathrm{ph}}-\delta^i_z\right)^2}{2\sigma_z^2}\right]\;,
\label{eq:p_t}
\end{equation}
where the photo-$z$ scatter $\sigma_z= 0.03(1+z)$  \citep{LSST18}, $\delta^i_z$ is the photo-z bias in the $i$-th tomographic redshift bin. And we assume $\delta_z = 0$ for the 4th generation surveys. Therefore, the reconstructed redshift distribution of SNIa in a given tomographic bin is \citep{Ma06,Wang23}
\begin{equation}
p^i(z) = \int_{z^i_{\rm min}}^{z^i_{\rm max}} dz_{\rm ph} p^i_{\rm ph}(z_{\rm ph}) p_{\rm t}(z \mid z_{\mathrm{ph}})\;,
\label{eq:p_z}
\end{equation}
where $p^i_{\rm ph}(z)$ is the distribution of photo-$z$ in the $i$-th tomographic redshift bin.

\begin{figure}
\plotone{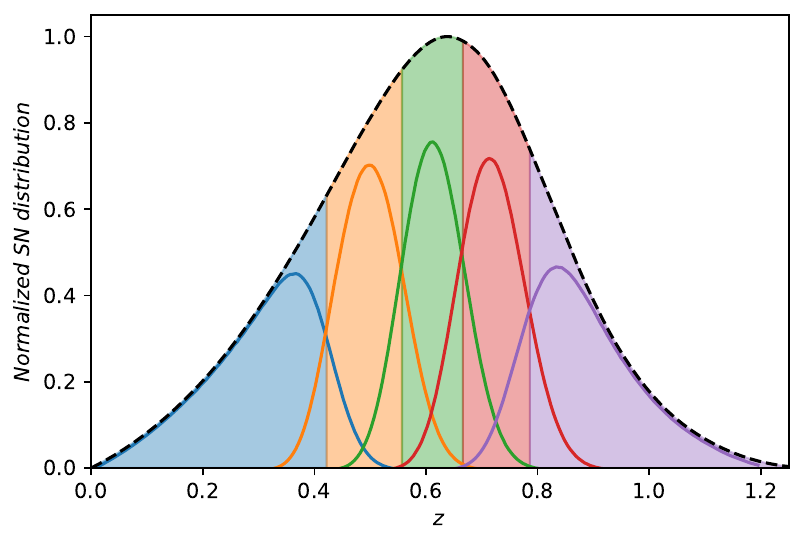}
\caption{ Mock SNIa redshift distributions in the WFD field of the LSST. The black dashed curve denotes the fiducial redshift distribution. The colored shaded areas indicate 5 tomographic redshift bins that contain the same number of SNIa. The solid colored curves are the true redshift distribution of the source whose photo-$z$ are measured within the tomographic redshift bins.}
\label{fig:z} 
\end{figure}

As shown in Fig. \ref{fig:z}, the black dashed line denotes the fiducial redshift distribution $p'\left(z\right)$, which is obtained based on SN detection efficiency and redshift distribution in the DC2-SNIa area \citep{Sanchez22}. Then, we divide the redshift range into five tomographic redshift bins (colored shaded areas) with the same number of SNIa. We utilized the reconstructed redshift distribution $p\left(z\right)$ obtained through Eq. \ref{eq:p_z} to estimate the true redshift distribution $p'\left(z\right)$ in each tomographic redshift bin, and showed them with colored solid lines in Fig. \ref{fig:z}. It can be found that the use of photo-$z$ to divide the tomographic redshift bins only allows a few sources at the edge to enter the adjacent tomographic redshift bin, so the data in each tomographic redshift bin can be used to construct its own magnification map.

Under a flat $\Lambda$CDM universe ($\Omega_M+\Omega_{\Lambda}=1$), the luminosity distance $D_{\mathrm{L}}$ can be written \citep{Kessler17}
\begin{equation}
D_{\mathrm{L}}\left(z, w, \Omega_{\mathrm{m}}\right) = (1+z) \frac{c}{H_0} \int_0^z \frac{d z^{\prime}}{E\left(z^{\prime}\right)}
\label{eq:dl}
\end{equation}
we define the function
\begin{equation}
E(z) = \left[\Omega_{\mathrm{m}}(1+z)^3+\Omega_{\Lambda}(1+z)^{3(1+w)}\right]^{1 / 2},
\end{equation}
where we assume the equation of state of dark energy $w=-1$. All the sources are randomly distributed on the sky map, and do not consider the clustering effect. Then, we calculate the fiducial luminosity distance $D_{\mathrm{L}}$ at a given redshift $z$ by Eq. \ref{eq:dl}. For all sources, the final catalog includes their fiducial redshift $z$, photo-$z$, fiducial luminosity distance $D_{\mathrm{L}}$ and positions in sky.

\subsection{Mock Supernovae Magnification Map}
\label{subsec:map}

The generation of the mock supernovae magnification map mainly consists of two steps. Firstly, we generate the theoretical magnification angular power spectrum within each tomographic redshift bin, and construct the magnification random realizations. Secondly, we generate the observed luminosity distance due to weak gravitational lensing effect, and reconstruct the magnification map by analyzing the observed luminosity distance of SNIa in WFD field. 

In the case of Gaussian statistics, it is easy to generate the realizations of magnification maps by standard technique. To satisfy both the auto- and cross-correlation of different each tomographic redshift bins, $\left\langle a^i_{\ell m} a^j_{\ell m}\right\rangle = C_{\ell}^{ij}$, we extend the method proposed by \cite{Kamionkowski97,Wang:2022uel} to study the realizations of temperature and polarization maps to multiple tomographic redshift bins, and correlate the spherical harmonic coefficients $a_{\ell m}$ between different  tomographic redshift bins with the same $\ell$ and $m$ using the following method,
\begin{eqnarray}
a_{\ell m}^{1} &=&\zeta_1\left(C_{\ell}^{11}\right)^{1 / 2}, \nonumber\\
a_{\ell m}^{2} &=&\zeta_1 \frac{C_{\ell}^{12}}{\left(a_{\ell m}^{1}/\zeta_1\right)^{1 / 2}}+\zeta_2\left[C_{\ell}^{22}-\frac{\left(C_{\ell}^{12}\right)^2}{a_{\ell m}^{1}/\zeta_1}\right]^{1 / 2}, ...\ , \nonumber\\
a_{\ell m}^{i} &=&\zeta_1 \frac{C_{\ell}^{1i}}{\left(a_{\ell m}^{1}/\zeta_1\right)^{1 / 2}}+...+\zeta_{i-1} \frac{C_{\ell}^{(i-1)i}}{\left(a_{\ell m}^{i-1}/\zeta_{i-1}\right)^{1 / 2}} \nonumber\\
&&+\zeta_i\left[C_{\ell}^{ii}-\frac{\left(C_{\ell}^{1i}\right)^2}{a_{\ell m}^{1}/\zeta_1}-...-\frac{\left(C_{\ell}^{(i-1)i}\right)^2}{a_{\ell m}^{i-1}/\zeta_{i-1}}\right]^{1 / 2}
\label{eq:alm_cross}
\end{eqnarray}

where for each given set $\ell$ and $m>0$, we choose a set of complex numbers $( \zeta_1, \zeta_2,...\ ,\zeta_i )$ drawn from a Gaussian distribution with unit variance; for $m = 0$, the $\zeta$ are real numbers drawn from a normal distribution; for $m < 0$, $a_{\ell m} = (-1)^{-m} a_{\ell, -m}$.

In order to ensure that the number of SNIa in each tomographic redshift bin is the same, the boundaries between the five tomographic redshift bins are set to $z=\{ 0.,\ 0.40,\ 0.54,\ 0.65,\ 0.79,\ 1.2\}$. Then we calculate the theoretical magnification angular power spectrum using Eq. \ref{eq:cluu}, with $\int p^i\left(z \right)dz =1$ in the $i$-th tomographic redshift bin, and produce mock magnification realizations from the theoretical angular power spectrum. The magnification image construction are based on the HEALPix software. In order to simulate the observational data as real as possible, we generate high resolution magnification realizations and set the HEALPix parameter $N_{\mathrm{side}} = 2048$, the pixel-scale $\sim 1.7$ arcmin pixel$^{-1}$. 
Since the magnification signal in the lowest redshift is weak, we only analyze the four high redshift bins. 

After generating the mock magnification images in each tomographic redshift bin, we use the Eq. \ref{eq:D_L} to obtain the lensed luminosity distance $D_{\mathrm{L}}^{\mathrm{len}}$, and add a Gaussian error of distance modulus as the mock observed luminosity distance $D_{\mathrm{L}}^{\mathrm{obs}}$. The distance modulus $\mu_{\rm DM}$ is defined by \citep{Holsclaw10}
\begin{equation}
\mu_{\rm DM} = m-M = 5\log\left(\frac{D_{\mathrm{L}}}{10 pc} \right)\;.
\end{equation}
So, the uncertainty in the observed luminosity distance $\sigma_{D_{\mathrm{L}}}$ can be calculated from the uncertainty in distance modulus $\sigma_{\mu_{\rm DM}}$ using the following
\begin{equation}
\sigma_{D_{\mathrm{L}}} = 0.2\mathrm{In}\left(10\right) D_{\mathrm{L}} \sigma_{\mu_{\rm DM}} \simeq 0.461D_{\mathrm{L}} \sigma_{\mu_{\rm DM}}
\end{equation}
where $\sigma_{\mu_{\rm DM}} =0.12$ is the typical value of SNIa intrinsic luminosity scatters \citep{Bernstein12,Scovacricchi17,LSST18,Arendse23}.

We divided all the SNIa into the tomographic redshift bins according to photo-$z$.

Then we can average all samples to obtain the mean luminosity distance $\overline{D}_{\mathrm{L}}$ for a redshift bin, and calculate the spatial magnification fluctuations by

\begin{equation}
\label{eq:mag_bin}
\delta_{\mu} = \left(\frac{D_{\mathrm{L}}^{\mathrm{obs}}}{\overline{D}_{\mathrm{L}}}\right)^2 -1\;.
\end{equation}

Due to the limitation of the total number of SNIa, we can obtain the magnification fluctuations maps with $N_{\mathrm{side}} = 128$, the number density $\sim 2/\mathrm{pixel}/\mathrm{bin}$, so we set the $\ell_{\rm max}$ for the magnification power spectrum to 383 ($3 \times N_{\mathrm{side}} -1$). The lower number density will cause some pixels to miss SNIa when randomly sprinkle dots, and we can't get any information about the magnification in these regions, so we need to remove these regions when calculating the power spectrum. The mask map of the tomographic redshift bin with $0.65<z_{\mathrm{photo}}\le1.2$ is showed in the right-hand panel of Fig. \ref{fig:map0}, we can find that about $8\%$ of the pixels in the whole WFD field are removed.

\section{galaxy shear}
\label{shear}

The CSST main sky survey areas will cover about 17,500 $\mathrm{deg^2}$ area with multi-band (from NUV to NIR) photometric imaging. It can perform photometric survey with high-spatial resolution (80\% energy concentration is $0.15''$) and deep limit magnitude (5$\sigma$ point source magnitude limit is 26 AB $\mathrm{mag}$ in $r$ band) \citep{Zhan11,Gong19,Cao22}. Hence, CSST will obtain excellent galaxy shape measurements, and is expected to provide smaller biases and sigma in the cosmological parameters constraints compared to stage-III surveys.

\subsection{Model of Galaxy Shear}
\label{subsec:shear_model} 

In tomographic weak lensing measurements, we need to consider the impact of residual additive and multiplicative systematics \citep{Amara08}, so the observed shear angular power spectrum for the $i$-th and $j$-th tomographic redshift bins can be written,
\begin{equation}
\hat{C}_{\gamma \gamma}^{ij}(\ell)=\left(1+m_i\right)\left(1+m_j\right) C_{\gamma\gamma}^{i j}(\ell)+N_{\mathrm{add}}^{i j}(\ell)+ \delta_{i j}N_{\mathrm{shot}}(\ell)\;,
\end{equation}
where the multiplicative systematic term dependent on the signal. 
We introduce a calibration parameter $m_i$ for the $i$-th tomographic redshift bin. ${C}_{\gamma \gamma}^{ij}(\ell)$ is the signal power spectrum, $N_{\mathrm{add}}^{i j}(\ell) = 10^{-9}$ is the additive systematic term with no redshift evolution, and shot-noise $N_{\mathrm{shot}}(\ell) = \sigma_\gamma^2/\bar{n}_i$. Here, the shear variance $\sigma_\gamma^2 = 0.09$ is taken from \cite{Gong19}, $\bar{n}_i$ is the average galaxy number density in the $i$-th tomographic redshift bin.

The galaxy shape is affected by weak gravitational lensing, and we can use galaxy shear to infer the weak gravitational lensing between galaxies and observers. However, shear measurement is contaminated by intrinsic alignment (IA), so the observed shape signal contains two components: the gravitational shear and intrinsic shape, $\gamma = \gamma^{\rm G} + \gamma^{\rm I}$. In this work, we adopt the "TATT" model including both tidal alignment (linear) and tidal torquing (quadratic) terms \citep{Blazek19}. Since the weak gravitational lensing and IA contributions to the B-mode power spectra are both very small, we only consider the influence of E-mode on the galaxy shear power spectrum. The total shear power spectrum is written \citep{Hirata04,Bridle07,Hikage19,Asgari21,Abbott22},
\begin{equation}
C_{\gamma\gamma}^{i j}(\ell)=C_{\mathrm{GG}}^{i j}(\ell)+C_{\mathrm{GI}}^{i j}(\ell)+C_{\mathrm{II}}^{i j}(\ell)\;.
\end{equation}
$C_{\mathrm{GG}}^{i j}(\ell)$ is the gravitational-gravitational power spectrum, which is the standard lensing signal and is identical to the convergence power spectrum. $C_{\mathrm{GI}}^{i j}(\ell)$ is the gravitational-intrinsic (GI) power spectrum, which denotes the correlation between the gravitational shear in one galaxy and the intrinsic shape of the other. $C_{\mathrm{II}}^{i j}(\ell)$ is the intrinsic-intrinsic (II) power spectrum, which denotes the correlation of the intrinsic shapes between neighboring galaxies. Under the Limber-approximate \citep{Limber53}, the above spectra between the $i$-th and $j$-th tomographic redshift bins are given by
\begin{eqnarray}
{C}^{ij}_{\rm GG}\left(\ell\right) &=& \int_{0}^{z_{\mathrm{H}}}dz\frac{d\chi}{dz}\frac{W_{\gamma}^{i}(z)W_{\gamma}^{j}(z)}{\chi^2}P_{\delta}(k=\ell/\chi,z),\\  {C}^{ij}_{\rm GI}\left(\ell\right) &=& \int_{0}^{z_{\mathrm{H}}}dz\frac{d\chi}{dz}\frac{W_{\gamma}^{i}(z)n^{j}(z)}{\chi^2}P_{\rm GI}(k=\ell/\chi,z)\nonumber\\
&+&\int_{0}^{z_{\mathrm{H}}}dz\frac{d\chi}{dz}\frac{n^{i}(z)W_{\gamma}^{j}(z)}{\chi^2}P_{\rm GI}(k=\ell/\chi,z)\;,
\end{eqnarray} 
and
\begin{equation}
{C}^{ij}_{\rm II}\left(\ell\right) = \int_{0}^{z_{\mathrm{H}}}dz\frac{d\chi}{dz}\frac{n^{i}(z)n^{j}(z)}{\chi^2}P_{\rm II}(k=\ell/\chi,z)\;,
\end{equation}
where $n\left(z \right)$ is the normalized distribution of galaxies from CSST and $W_{\gamma}^{i}(z)$ is the lensing weighting function for the CSST tomographic redshift bins. It is written as
\begin{equation}
W_{\gamma}^i(z) =\frac{3 \Omega_m H_0^2 \chi}{2c^2 a} \int_{z}^{z_{\mathrm{H}}}d\chi' \frac{d\chi'}{dz}n^i\left(z' \right)\frac{\chi'-\chi}{\chi'}\;.
\end{equation}

The GI and II power spectra can be expressed by the following
\begin{eqnarray} 
P_{\rm GI}\left(k\right)&=&\ C_1  P_\delta(k)+C_{1 \delta}\left[A_{0 \mid 0 E}\left(k\right)+C_{0 \mid 0 E}\left(k\right)\right] \nonumber\\
&&+ C_2\left[A_{0 \mid E 2}\left(k\right)+B_{0 \mid E 2}\left(k\right)\right]\;, \nonumber\\
P_{\rm II}\left(k\right)&=&\ C_1^2  P_\delta(k)+2 C_1 C_{1 \delta} \left[A_{0 \mid 0 E}\left(k\right)+C_{0 \mid 0 E}\left(k\right)\right] \nonumber\\
&&+ C_{1 \delta}^2 A_{0 E \mid 0 E}\left(k\right) +C_2^2 A_{E 2 \mid E 2}\left(k\right) \nonumber\\
&&+2 C_1 C_2 \left[A_{0 \mid E 2}\left(k\right)+B_{0 \mid E 2}\left(k\right)\right] \nonumber\\
&&+2 C_{1 \delta} C_2 D_{0 E \mid E 2}\left(k\right)\;.
\end{eqnarray} 
These $k$-dependent terms are discussed in \cite{Blazek19}, we evaluate them using the FAST-PT module \citep{Fang17} of Core Cosmology Library\footnote{\url{https://github.com/LSSTDESC/CCL}} (CCL) \citep{Chisari19}. $C_{1 \delta}(z)$ is the linear galaxy biasing, it is generally thought to come purely from density weighting effect, $C_{1 \delta}(z) = b_1 C_1(z)$.
For simplicity, we fix $b_1$=0 here. $C_1(z)$ and $C_2(z)$ can be written as
\begin{eqnarray}
C_1(z)&=& -c_1 \bar{C}_1 \rho_{\text {crit}} \frac{\Omega_{\mathrm{m}}}{D(z)}\;,\nonumber\\
C_2(z)&=& 5c_2 \bar{C}_1 \rho_{\text {crit}} \frac{\Omega_{\mathrm{m}}}{D^2(z)}\;,
\end{eqnarray}
where the amplitudes $c_1$ and $c_2$ are two free parameter, which are assumed as $c_1$= $c_2$ = 1.
The empirical amplitude reads $\bar{C}_1= 5 \times 10^{-14} h^{-2}M_{\odot}^{-1}\mathrm{Mpc^3}$ \citep{Brown02}, $\rho_{\text {crit}}$ is the present critical density, ${D(z)}$ is the linear growth factor, normalized to unity at $z=0$.

\begin{figure}
\plotone{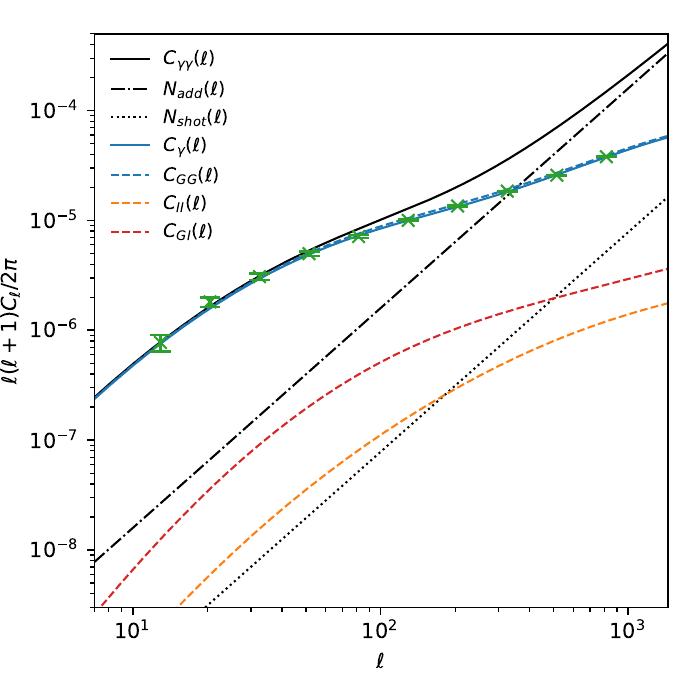}
\caption{Components of the galaxy shear power spectrum for the tomographic redshift bin with $0.6<z_{\mathrm{photo}}\le1.2$. The black solid line and dashed line denote the total power spectrum with noise and the power spectrum of the signal, respectively. The additive systematic noise and shot-noise are shown in black dashed-dotted and dotted lines, respectively. We assume the additive systematic noise $N_{\mathrm{add}}(\ell) = 10^{-9}$ for the CSST survey, and assume the IA include both tidal alignment and tidal torquing terms. The blue, orange and red dashed lines represent the gravitational-gravitational, intrinsic–intrinsic and gravitational-intrinsic power spectra, respectively. Note that $C_{\rm GI}(\ell)$ is negative value. For convenience of display, we take its absolute value as the red dashed line.}
\label{fig:TATT} 
\end{figure}

In Fig. \ref{fig:TATT}, we show the power spectra discussed above for the tomographic redshift bin with $0.6<z_{\mathrm{photo}}\le1.2$. The black solid line and dashed line denote the total power spectrum with noise and the power spectrum of the signal, respectively. 
The additive systematic noise and shot-noise are shown with black dashed-dotted and dotted lines, respectively. The blue, orange and red dashed curves are the power spectra of GG, II and GI, respectively.

\subsection{Mock Shear Map}
\begin{figure}
\plotone{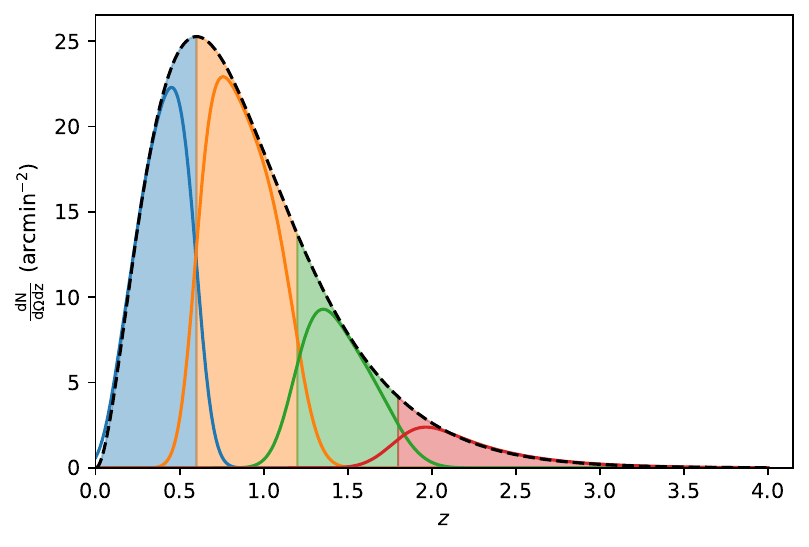}
\caption{ The redshift distributions of the CSST. We divide the redshift range into several four photo-$z$ bins for tomographic analysis, which are shown as the colored shaded areas. The solid color lines indicate the true redshift distribution of the source whose photo-$z$ are measured within the tomographic redshift bins.}
\label{fig:z_shear} 
\end{figure}

Following \cite{Miao23}, we adopt a galaxy redshift distribution $n(z) \propto z^{\alpha}e^{-(z/z_0)^{\beta}}$, which obtained by fitting the COSMOS catalog for the CSST survey \citep{Capak07,Ilbert09}. We set $\alpha = 2$, $\beta = 1$ and $z_0 = 0.3$, and assume a total number density $n = 28\ \rm{arcmin}^{-2}$ based on the CSST astrometric capability. The redshift distribution of the galaxy samples, as well as our binning scheme, is shown in Fig. \ref{fig:z_shear}. We can find that the redshift distribution of the total galaxy samples has a peak around $z = 0.6$, and can extend to $z \sim 4$. Similar to the treatment of SNIa, we also divide the redshift range into several different photo-$z$ bins for tomographic analysis. \cite{Gong19} discussed the relationship between the number of tomographic redshift bins and constraint results on cosmological parameters. 
We adopt four tomographic redshift bins as those in \cite{Gong19}. 
The first three tomographic bins have equal redshift intervals $\Delta_z = 0.6$, and the last tomographic bin occupies the remaining range of redshift distribution. In Fig. \ref{fig:z_shear}, the photo-$z$ bins are represented by colored shaded areas, the total redshift distribution $n\left(z\right)$ is represented by the black dashed line. We adopt the photo-$z$ scatter $\sigma_z= 0.05(1+z)$ and photo-$z$ bias $\delta_z = 0$ as the fiducial values for the CSST survey \citep{Cao18}. Then, we estimate the true redshift distribution of the galaxy samples within the given tomographic photo-$z$ bin by the Eq. \ref{eq:p_t} and Eq. \ref{eq:p_z}, and show them as the colored solid curves.

\begin{figure*}
\gridline{\fig{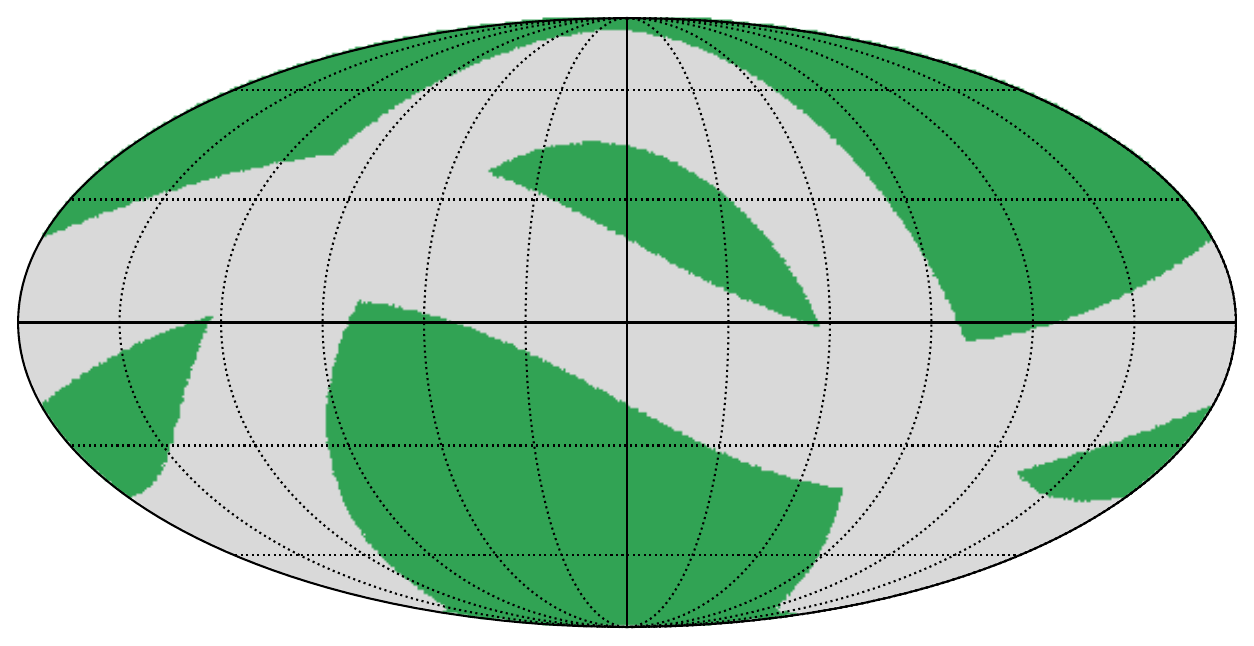}{0.49\textwidth}{}
          \fig{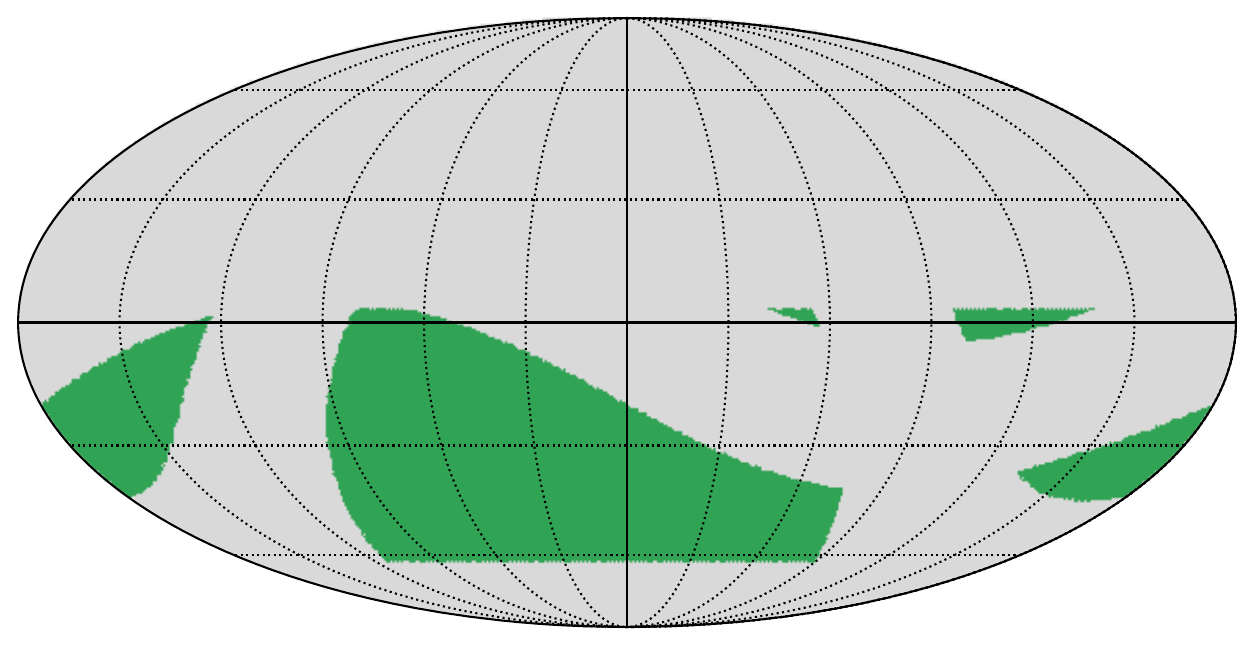}{0.49\textwidth}{}}
\caption{ $\it left$: The distribution of the galaxy samples for CSST cosmic shear forecast. The green area is the main sky survey field, which cover the median-to-high Galactic latitude ($|b|>19.2^{\circ}$) and median-to-high Ecliptic latitude ($|\beta|>19.2^{\circ}$) of the sky, which cover about 17,500 $\rm deg^2$ area and is expected to allocate $\sim70\%$ of the total operation time. $\it right$: The overlap areas of CSST and LSST-WFD surveys. these areas cover about $7,000 {\rm deg^2}$.}
\label{fig:field_CSST} 
\end{figure*}

The observation strategy of CSST is based on operational scheduling constraints and observation requirements. In order to meet the demands of cosmological studies and reduce the influence of stellar and zodiacal light background, CSST is required to observe in the median-to-high Galactic latitude and median-to-high Ecliptic latitude of the sky. This will allow it to obtain a large sample of extragalactic sources \citep{Zhan21}. We show the CSST fields in the left-hand panel of Fig. \ref{fig:field_CSST}, and the sky is shown in HEALPix projection in equatorial over a 10-year time span. The green area is the main sky survey field, which cover about 17,500 $\rm deg^2$ area and is expected to allocate $\sim 70\%$ of the total operation time. For a given survey schedule, we remove the regions within $\pm19.2^{\circ}$ of the Galactic latitude and the Ecliptic latitude of the sky to generate the simulated shear map \citep{Fu23,Yao23}. 

We first obtain the auto- and cross-power power spectra of the SNIa magnification and galaxy shear in each tomographic redshift bin, and convert the these spectra into the spherical harmonic coefficients $a_{\ell m}$ by the Eq \ref{eq:alm_cross}, and generate some HEALPix maps with $N_{\mathrm{side}} = 4096$ (the pixel-scale $\sim 0.86$ arcmin pixel$^{-1}$) as the fiducial shear maps of CSST survey. Then, we add the systematic niose and shot-niose to the fiducial shear maps as simulated observations. Here, the systematic niose contains an additive term with no redshift evolution and a multiplicative term. 
The additive bias is added into maps and multiplicative bias is multiplied in the power spectrum.   
Finally, we remove the regions outside the CSST target sky.

\subsection{Cross-correlation of Galaxy Shear and SNIa Magnification}
\label{subsec:cross_ru}

Considering the impact from shape measurement systematics and IA, the cross angular power spectrum of the $i$-th LSST tomographic redshift bin and $j$-th CSST tomographic redshift bin can be written
\begin{equation}
\hat{C}_{\mu \gamma}^{i j}(\ell)=\left(1+m_j\right)\left[ C_{\mathrm{\mu G}}^{i j}(\ell)+C_{\mathrm{\mu I}}^{i j}(\ell) \right]\;.
\end{equation}
Here, the magnification-shear power spectrum $C_{\mathrm{\mu G}}^{i j}(\ell)$ is estimated by the cross-correlation of the magnification and convergence. Magnification-intrinsic power spectrum $C_{\mathrm{\mu I}}^{i j}(\ell)$ denotes the 
correlation between the gravitational lensing of foreground and the intrinsic ellipticity of background galaxy. Under the Limber-approximation, the cross spectrua are given by
\begin{equation}
{C}^{ij}_{\rm \mu G}\left(\ell\right) = \int_{0}^{z_{\mathrm{H}}}dz\frac{d\chi}{dz}\frac{W_{\mu}^{i}(z)W_{\gamma}^{j}(z)}{\chi^2}P_{\delta}(k=\ell/\chi,z),
\end{equation}
and
\begin{equation}
{C}^{ij}_{\rm \mu I}\left(\ell\right) = \int_{0}^{z_{\mathrm{H}}}dz\frac{d\chi}{dz}\frac{W_{\mu}^{i}(z)n^{j}(z)}{\chi^2}P_{\rm \mu I}(k=\ell/\chi,z)\;.\\
\end{equation}
Here, the $\rm \mu I$ power spectra $P_{\rm \mu I} = P_{\rm GI}$.

In order to calculate the cross-power spectra of CSST shear maps and LSST SNIa magnification maps, we select the overlap area that cover about $7,000 {\rm deg^2}$ as the target region, and show the area in the right-hand panel of Fig. \ref{fig:field_CSST}. The CSST shear maps are made at a pixel-scale of 0.86 arcmin pixel$^{-1}$. Furthermore, we need smooth the maps to a 27 arcsec pixel$^{-1}$ scale to match the pixel-scale of our magnification maps.

\section{measurements of power spectra}
\label{spectrum}
The auto- and cross-spectra of the HEALPix maps are estimated using the technique of spherical harmonics transformation. However, the power spectra are affected by statistical measurements, masked areas and noise. In this section, we describe the spectrum correction and covariance analysis.

\subsection{Correction of Power Spectrum}
\label{Cl_correction}

The raw angular power spectra $\widetilde{C}_{\ell}$ obtained by direct measurement of masked map is contaminated by some pseudo signals. In this work, we correct the raw $\widetilde{C}_{\ell}$ using the following 
\begin{equation}
\widetilde{C}_{\ell}=M_{\ell \ell^{\prime}} C_{\ell^{\prime}} + N_{\ell},
\end{equation}
where $C_{\ell^{\prime}}$ is the true angular power spectrum from full sky. $M_{\ell \ell^{\prime}}$ is the mode coupling matrix, which can correct the errors introduced by the mask, and $N_{\ell}$ is the noise.

In actual observations, due to insufficient sampling, there may be some areas that do not contain any observed data. So it is necessary to remove these areas in the analysis of intensity fluctuations. However, using a mask will introduce additional pseudo signal in the estimation of the power spectrum. For a mask with a very simple shape, the profile change of the power spectrum due to the mask can be ignored, so the raw power spectrum is easily corrected by dividing the coefficient $f_{\rm sky}$ which is the fraction of the unmasked areas of the whole sky. For complex-shaped mask, the mask can introduce a leakage from the large $\ell$ modes into small ones, so that the profile of the power spectrum will change significantly. In this section, we generate the mode coupling matrix $M_{\ell \ell^{\prime}}$ by analyzing the effects of the mask at $\ell$-mode, and use it to correct the changes of the power spectrum \citep{Cooray12,Zemcov14}.

\begin{figure}
\plotone{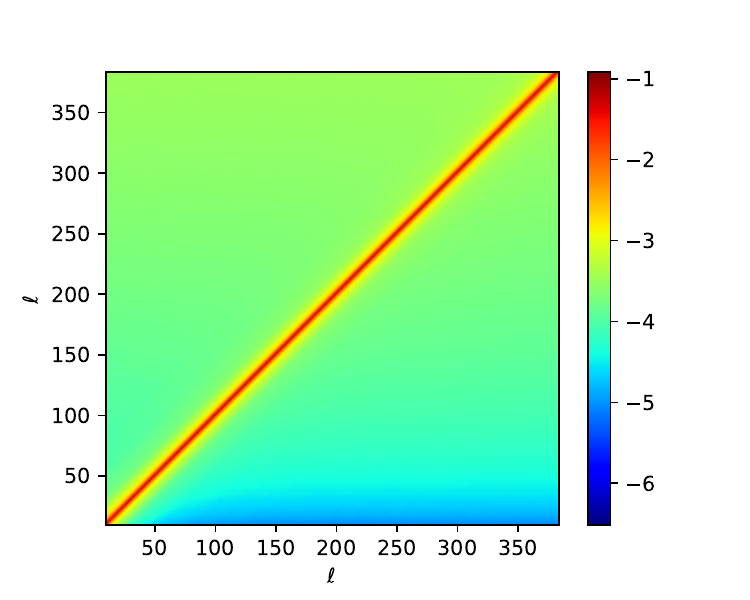}
\caption{The mode coupling matrix $M_{\ell\ell^{\prime}}$ for the SNIa magnification map. The $\ell$ modes cover from $\ell = 30$ to $\ell = 383$, and the coordinates of the color bar is logarithmic scale. We can find that the effect between adjacent bins is greater, and high $\ell$ modes (small scale) has a greater effect than low $\ell$ modes (large scale). }
\label{fig:mll}
\end{figure}

For a given mask, we calculate the mode coupling matrix $M_{\ell \ell^{\prime}}$ by a simulation procedure. First, we construct a realization of map from a pure tone power spectrum (where $C_{\ell^{\prime}} = 1$ if $\ell^\prime=\ell$, otherwise $C_{\ell^{\prime}} = 0$), and remove the pixels using the given mask. Second, we calculate the raw power spectrum $\widetilde{C}_{\ell^{\prime}}$ of the masked maps, and use the result show how the mask mixes the power spectrum from $\ell-$mode into other modes. Third, we repeat the previous two steps 100 times and take the average of all the simulation results as the corresponding row for $\ell$ in the mode coupling matrix $M_{\ell \ell^{\prime}}$. It is can be expressed as $M_{\ell}=\left\langle{\widetilde{C}}_{\ell^{\prime}}\right\rangle$. Finally, we repeat the entire process mentioned above for all other $\ell$, and obtain the mode coupling matrix $M_{\ell \ell^{\prime}}$. After this, we calculate the inverse of the mode coupling matrix to correct for the effect from the mask by $C_{\ell^{\prime}}= M_{\ell \ell^{\prime}}^{-1}\widetilde{C}_{\ell}$. We show an example of the mode-coupling matrix of a SNIa magnification map in Fig. \ref{fig:mll}. 
 
Generally, a map is thought to consist of signal $S$ and noise $N$. For two different detectors or imaging experiments, the cross-correlation of different maps can be expressed as $M_1 \times M_2 = (S_1 +N_1)\times(S_2 +N_2)$. Since noise is random, one noise should be uncorrelated with another noise or signal, so the cross-correlation can minimize the contamination from instrumental noise \citep{Thacker15,Cao20}. In this work, we first estimate the power spectrum of the imaging data generated from all sample sources. Then, we randomly divide all sample sources into two groups and generate the cross-power spectrum of the two mapping images. Finally, we take the difference between the auto- and cross-correlation as noise, and use a Gaussian distributed white noise $N_{\ell}$ for fitting.

\subsection{Covariances}
\label{Covariances}

For simplicity, we only consider the Gaussian covariance to account for the statistical error for the imaging data. The covariance matrix of the power spectra can be estimated by \citep{Hu04,Huterer06,Joachimi08}

\begin{eqnarray}
{\rm Cov}_{ijmn}\left[\hat{C}_{XY}^{ij}(\ell),\hat{C}_{X^{\prime}Y^{\prime}}^{mn}(\ell')\right]= \frac{\delta_{\ell \ell'}}{f_{\rm sky}\Delta \ell(2\ell+1)} \nonumber\\
\times\left[\hat{C}_{XX^{\prime}}^{im}(\ell)\hat{C}_{YY^{\prime}}^{jn}(\ell)+\hat{C}_{XY^{\prime}}^{in}(\ell)\hat{C}_{YX^{\prime}}^{jm}(\ell)\right]&,
\end{eqnarray}

where $X$, $Y$, $X^{\prime}$ and $Y^{\prime}$ are either the SNIa magnification $\mu$ or the galaxy shear $\gamma$. Note that if auto-spectra on the right-hand side of the equation, they include all noise components. Limited by the number of SNIa samples, we estimate the magnification-magnification $\hat{C}_{\mu\mu}(\ell)$ and magnification-shear $\hat{C}_{\mu \gamma}(\ell)$ power spectra between $\ell_{\rm min} = 30$, to avoid deviation from the Limber approximation, and $\ell_{\rm max}=383$, to be limited by the SNIa number density. The angular frequency is smaller than the coverage of the shear-shear power spectra $\hat{C}_{\gamma \gamma}$, whose small-scale extends to $\ell_{\rm max}=3000$ (a conservative scale cut with $\ell_{\rm max}=1000$) to avoid using incorrect physical models on small scales. We set 15 logarithmic $\ell$ bins for $\hat{C}_{\gamma \gamma}$. However, the scales of $\hat{C}_{\mu \mu}$ and $\hat{C}_{\mu \gamma}$ cover a relatively small range, considering the signal-to-noise ratios (S/N) at both large-scale and small-scale, we have adopted 10 logarithmic $\ell$ bins. The sky coverage fraction $f_{\rm sky}$ varies for different power spectra. The LSST-WFD can cover $\sim16,000\ {\rm deg^2}$ after removing masked area, the CSST main sky survey areas can cover $\sim17,500\ {\rm deg^2}$, and their overlap is greater than $7,000\ {\rm deg^2}$.

\section{Results and analysis of power spectrum}
\label{spec_result}

Based on the survey capability, the auto- and cross-spectra of SNIa magnification and galaxy shear can be obtain after spherical harmonics transforming and correcting. In this section, we present the final power spectra and their uncertainty, and compare them with the fiducial value. Then, we use the theoretical model to fit these power spectra, and constrain the model parameters by performing a MCMC analysis. 

\subsection{Final Power Spectra}
\label{subsec:final_power}
\begin{figure*}
\centering
\includegraphics[width=2.0\columnwidth]{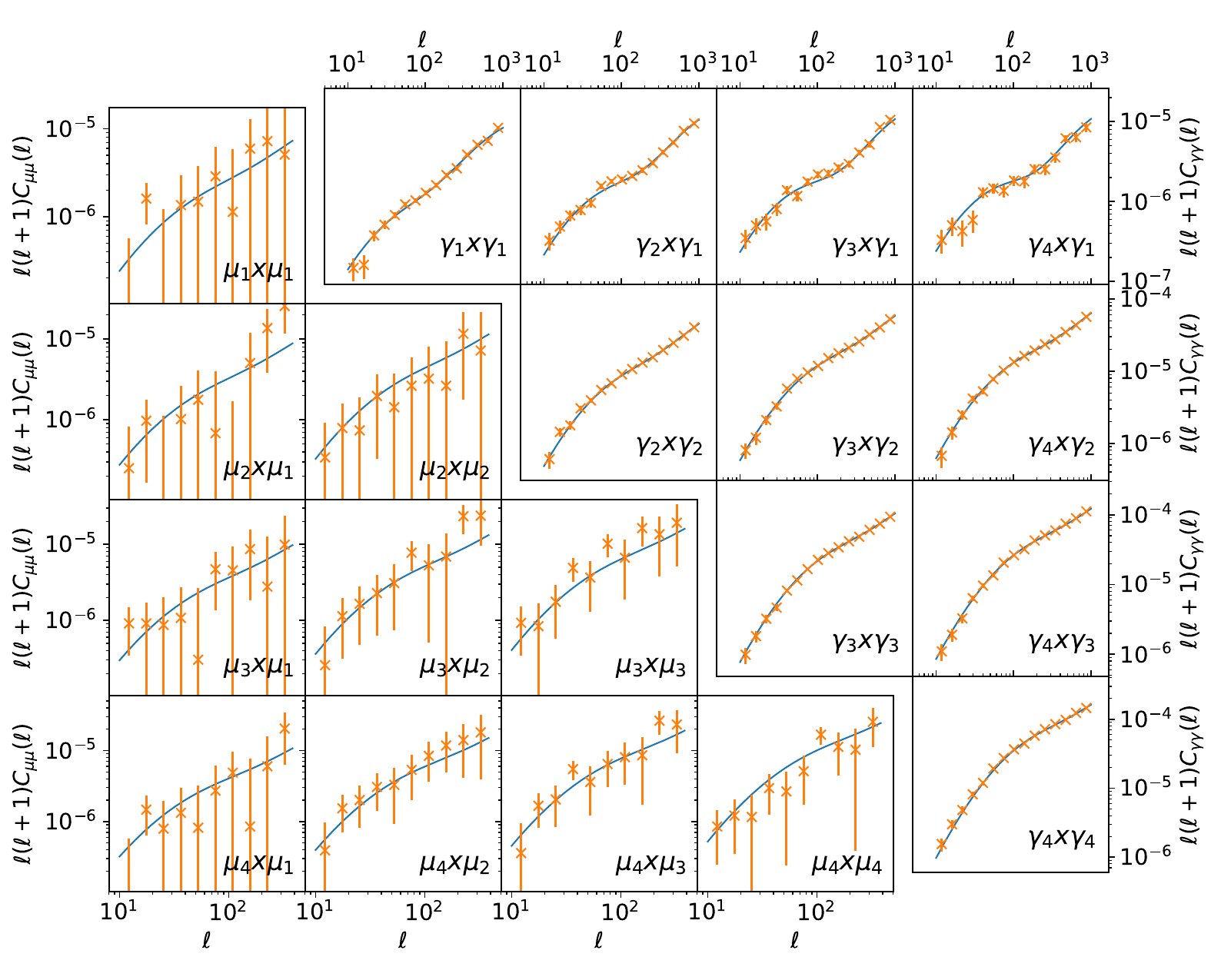}
\caption{ Auto- and cross-power spectra of the magnification-magnification $C_{\mu\mu}(\ell)$ and shear-shear $C_{\gamma \gamma}(\ell)$. We show the measured final power spectra subtracting the additive systematic and shot-noise in orange data points. The error bars are 1$\sigma$ uncertainties which are obtained by calculating the covariance matrix. The blue lines are the fiducial theoretical power spectrum.}
\label{fig:llcl}
\end{figure*}

\begin{figure*}
\centering
\includegraphics[width=2.0\columnwidth]{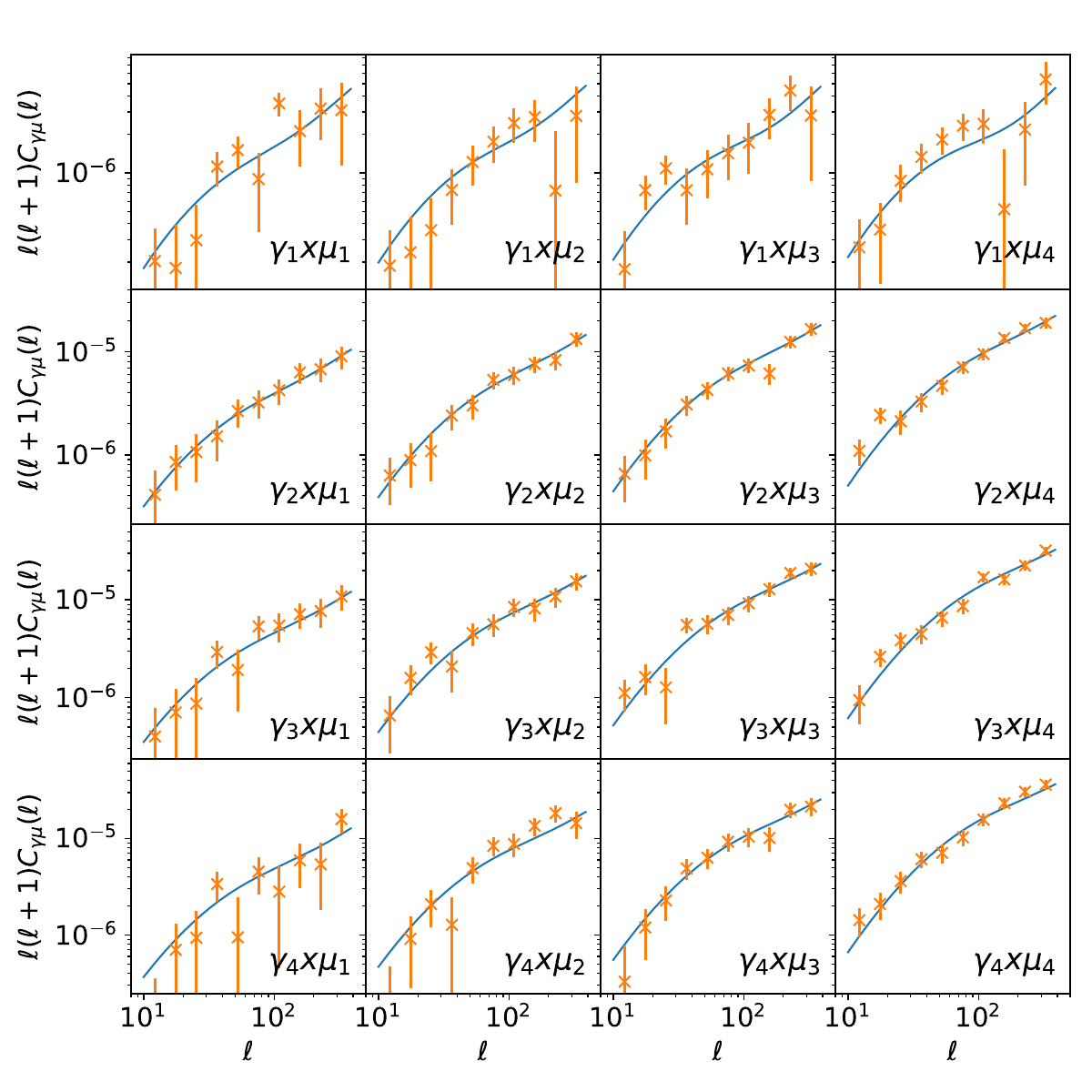}
\caption{The tomographic magnification-shear angular power spectra. In each subplot we show the cross-power spectra $C_{\mu\gamma}^{ni}(\ell)$ between the $n$-th LSST magnification tomographic bin and $i$-th CSST galaxy shear bin.}
\label{fig:llclur}
\end{figure*}

We estimate the auto- and cross-spectra of the simulated imaging data from the CSST and LSST surveys. The mask map removes the areas that do not contain measured data, and also remove the non-overlapping regions of CSST and LSST-WFD for the cross-correlation of magnification and shear. Then, the raw spectra are obtained from the mock maps, and they are corrected by the mode coupling matrix and transfer function. Allowing the shot-noise and additive systematic noise as free parameters in each tomographic bin, and eliminating them before fitting process can lead to improved constraints on cosmological parameters. After these operations, we show the final angular power spectra as $\ell(\ell+1)C_{\ell}/2\pi$ in Fig. \ref{fig:llcl} and Fig. \ref{fig:llclur}.

In Fig. \ref{fig:llcl}, we show the magnification-magnification $C_{\mu\mu}(\ell)$ and shear-shear $C_{\gamma \gamma}(\ell)$ power spectra in the bottom left and top right, respectively. Here, the corrected power spectra are presented as orange data points over cover from $\ell=$30 to 383, extend to $\ell=3000$ (a conservative scale $\ell=1000$)  for $C_{\gamma \gamma}$. The plotted error bars are 1$\sigma$ uncertainties, which are derived from the covariance matrix. We use  the blue curves to show the fiducial theoretical power spectrum. As can be seen, the S/N of the shear-shear $C_{\gamma\gamma}(\ell)$ is significantly higher than that of the magnification-magnification $C_{\mu\mu}(\ell)$. For the shear-shear $C_{\gamma\gamma}(\ell)$, we found that the S/N of cross-spectra between different tomographic bins decrease compared to the auto-spectra, especially for the two tomographic bins with relatively distant redshift. For the magnification-magnification $C_{\mu\mu}(\ell)$, the power spectra obtained from high-redshift samples have a higher S/N compared to those obtained from low-redshift samples due to the increase in signal strength. Cross-correlation can minimize the contamination from noise, allowing us to obtain signals from noise-dominated data, but the noise still increases the uncertainty of the cross-correlation power spectrum. The high uncertainty of the distance modulus increases the error of luminosity distance, it brings high noise to the SNIa magnification imaging data. Reducing the uncertainty of the distance modulus and increasing the sample number of SNIa can both effectively reduce the noise. We hope that future observations can obtain more sources with high S/N, which will help to measure the magnification. In Fig. \ref{fig:llclur}, we show the tomographic magnification-shear cross-power spectra $C_{\mu\gamma}^{ij}(\ell)$ between the $i$-th LSST SNIa magnification tomographic bin and $j$-th CSST galaxy shear bin. Although S/N of shear-shear measurement is high, it will inevitably be affected by the multiplicative systematic bias. When fitting the galaxy shear data, we need add more parameters in the model, this could lead to looser constraints. One advantage of SNIa magnification measurement is that it is not affected by multiplicative systematic bias. The joint of SNIa magnification measurement and galaxy shear can help improve the constraint capabilities and reduces degeneracies form the multiplicative systematic.

\subsection{Fitting results}
\label{subsec:fit_result}

\begin{figure*}
\centering
\includegraphics[width=2.\columnwidth]{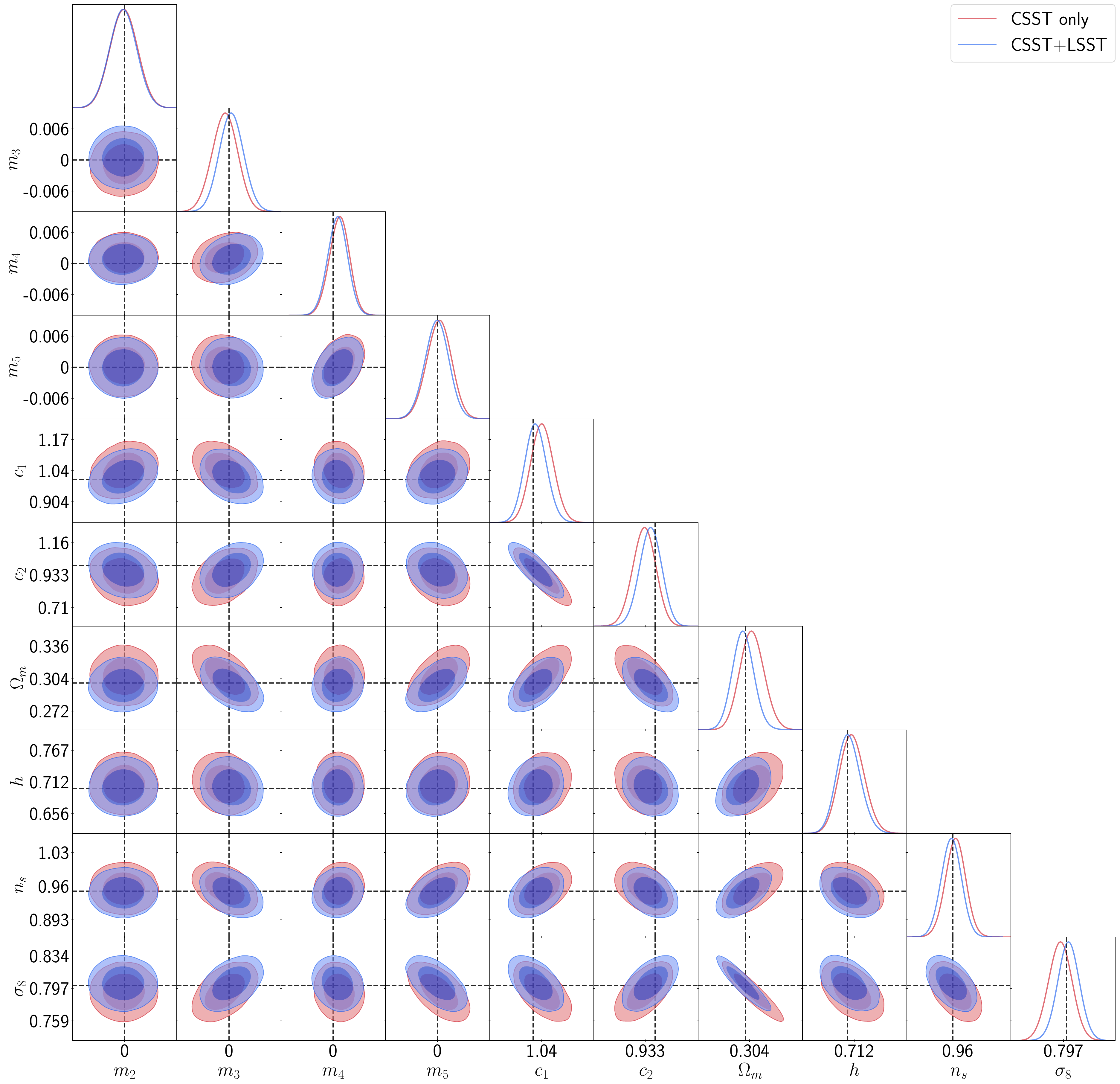}
\caption{ The contour maps with 1$\sigma$ (68.3\%) and 2$\sigma$ (95.5\%) C.L. of the free parameters, i.e. $\Omega_m$, $h$, $n_s$, $\sigma_8$, $c_1$, $c_2$ and four multiplicative calibration parameters $m$ under the condition of $\ell_{\rm max}=1000$ cut. Please note that due to the magnification signal in the lowest redshift is weak, we removed the for SNIa we have removed the smallest redshift bin for SNIa analysis. The colored solid curves are and the black dashed lines are the the 1-D PDFs and fiducial values of the parameters, respectively. The red contours denote the results of the galaxy shear survey, and the blue contours are the joint constraint results.}
\label{fig:fit} 
\end{figure*}

\begin{table*}
\centering
\caption{\label{tab:result} The best-fits and errors of the free parameters in the model from the MCMC fitting.}
\begin{tabular}{c c c c c c c}
\hline\hline
\rule[-1mm]{0mm}{4mm}
&&\multicolumn{2}{c}{$\rm Best\ Fit\ (\ell_{\rm max}=1000)$} &  \multicolumn{2}{c}{$\rm Best\ Fit\ (\ell_{\rm max}=3000)$} \\
\cline{3-4} \cline{5-6}
Parameter & $\rm Fiducial\ Value$ & $\rm Shear\ Only$ & $\rm Joint$ & $\rm Shear\ Only$ & $\rm Joint$ & $\rm Prior\ ( min,max)$\\
\hline
\multicolumn{7}{c}{Cosmology}\\
\rule[-2mm]{0mm}{6mm}
$\Omega_{\rm m}$      &$0.30$ &$0.307^{+0.011}_{-0.012}$  &$0.298^{+0.010}_{-0.011}$ &$0.303^{+0.009}_{-0.009}$  &$0.301^{+0.008}_{-0.008}$ &$0, 1$\\
\rule[-2mm]{0mm}{6mm}
$h$  &$0.7$ &$0.708^{+0.021}_{-0.023}$ &$0.702^{+0.020}_{-0.022}$ &$0.705^{+0.019}_{-0.020}$ &$0.698^{+0.018}_{-0.019}$ &$0, 1$\\
\rule[-2mm]{0mm}{6mm}
$n_{\rm s}$      &$0.95$ &$0.955^{+0.021}_{-0.021}$& $0.947^{+0.020}_{-0.020}$ &$0.953^{+0.014}_{-0.014}$& $0.950^{+0.013}_{-0.013}$&$\ \ 0.85, 1.1$\\
\rule[-2mm]{0mm}{6mm}
$\sigma_8$   &$0.8$ &$0.793^{+0.014}_{-0.014}$ &$0.802^{+0.013}_{-0.013}$ &$0.796^{+0.010}_{-0.010}$ &$0.798^{+0.009}_{-0.009}$ &$0.4, 1$\\
\hline
\multicolumn{7}{c}{Intrinsic alignment}\\
\rule[-2mm]{0mm}{6mm}
$c_1$                &$1.0$  &$1.039^{+0.049}_{-0.049}$ &$1.012^{+0.046}_{-0.047}$ &$1.031^{+0.031}_{-0.031}$ &$1.015^{+0.029}_{-0.030}$ &$0, 2$\\
\rule[-2mm]{0mm}{6mm}
$c_2$                &$1.0$  &$0.926^{+0.081}_{-0.077}$ &$0.974^{+0.078}_{-0.073}$   &$0.935^{+0.055}_{-0.054}$ &$0.958^{+0.053}_{-0.052}$ &$0, 2$\\
\hline
\multicolumn{7}{c}{Multiplicative systematic}\\
\rule[-2mm]{0mm}{6mm}
$m_2$                &$0.$&$-0.0002^{+0.0026}_{-0.0026}$ &$-0.0002^{+0.0026}_{-0.0026}$ &$0.0004^{+0.0026}_{-0.0026}$ &$0.0003^{+0.0026}_{-0.0026}$ &$-1, 1$\\
\rule[-2mm]{0mm}{6mm}
$m_3$                &$0.$&$-0.0008^{+0.0025}_{-0.0025}$  &$0.0005^{+0.0024}_{-0.0024}$ &$-0.0013^{+0.0023}_{-0.0022}$ &$-0.0009^{+0.0022}_{-0.0021}$ &$-1, 1$\\
\rule[-2mm]{0mm}{6mm}
$m_4$                &$0.$&$0.0012^{+0.0020}_{-0.0020}$ &$0.0008^{+0.0019}_{-0.0019}$ &$-0.0002^{+0.0017}_{-0.0017}$ &$0.0000^{+0.0017}_{-0.0017}$ &$-1, 1$\\
\rule[-2mm]{0mm}{6mm}
$m_5$                &$0.$&$0.0003^{+0.0024}_{-0.0024}$ &$-0.0001^{+0.0023}_{-0.0023}$ &$0.0013^{+0.0020}_{-0.0020}$ &$0.0012^{+0.0020}_{-0.0020}$ &$-1, 1$\\
\hline
\end{tabular}
\end{table*}

In order to explore the joint constraint of the SNIa magnification and galaxy shear on the cosmological parameter, we adopt the MCMC method \citep{Metropolis53} to fit the final power spectra using the theoretical model. We use the Monte Python software \citep{Audren13} to estimate the acceptance probability of the MCMC chain point, and to analyze the free parameters. The likelihood function can be estimate by $\mathcal{L} \sim \exp(-\chi^2/2)$, and the $\chi ^2$ distribution is calculated by
\begin{equation} \label{eq:chi2}
\chi^2=\sum_{ijmn}^{N} \left(C_{\rm obs}^{ij}-C_{\rm th}^{ij}\right){\rm Cov}_{ijmn}^{-1} \left(C_{\rm obs}^{mn}-C_{\rm th}^{mn}\right),
\end{equation}
where $N$ is the number of tomographic redshift bins. $C_{\rm obs}$ and $C_{\rm th}$ are the observed and theoretical power spectra, respectively. The inverse of the covariance ${\rm Cov}_{ijmn}^{-1}$ can obtain by the covariance matrix. The total $\chi_{\rm tot}^2$ for joint surveys of the LSST SNIa magnification and CSST galaxy shear is given by 
\begin{equation}
\chi_{\rm tot}^2 = \chi_{\mu\mu\mu\mu}^2 + \chi_{\gamma\gamma\gamma\gamma} ^2 + \chi_{\mu\gamma\mu\gamma} ^2 + \chi_{\mu\mu\mu\gamma} ^2 + \chi_{\mu\mu\gamma\gamma} ^2 + \chi_{\mu\gamma\gamma\gamma} ^2\;.
\end{equation}

The theoretical model contains 10 free parameters, including four cosmological parameters: $\Omega_m$, $h$, $n_s$ and $\sigma_8$; two IA model parameters: $c_1$ and $c_2$; and four multiplicative calibration parameters $m_i$, please note that due to the magnification signal in the lowest redshift is weak, we removed the for SNIa we have removed the smallest redshift bin for SNIa analysis. In this work, we set flat priors:  $c_1 \in ( 0,2)$, $c_2 \in ( 0,2)$, $\Omega_{\rm m} \in ( 0, 1)$, $h \in ( 0, 1)$, $n_{\rm s} \in$ (0.85, 1.1) and $\sigma_8 \in ( 0.4, 1)$. For the multiplicative parameters $m_i$, we adopt a Gaussian prior $\mathcal{N}(0,0.0026^2)$. $\sigma_m \le 0.01$ has been achieved in the stage-III surveys, such as DES \citep{Abbott22}, KiDS \citep{Hildebrandt17}, HSC \citep{Hikage19}, and the stage-IV surveys such as CSST, LSST and Euclid can achieve better accuracy. CSST requires that the systematic uncertainty in multiplicative calibration does not exceed 0.0026 \citep{Yao23}, which is similar to the 0.003 requirement of LSST \citep{LSST18}.
For each case, we run 20 chains, and each chain contains 100,000 steps. We have shown the fitting results and 68$\%$ confidence levels (C.L.) in Table~\ref{tab:result}.

There are 10 magnification-magnification power spectra and 16 magnification-shear power spectra, they all consist of 10 data points. The shear-shear power spectra consist 15 data points, so the degree of freedom is $N_{\rm dof} = 15\times10-10= 140$ for shear measurement, and $N_{\rm dof} = 140 +10\times26= 400$ for joint fitting. We find that the reduced chi-square $\chi^2_{\rm red}=\chi^2_{\rm min}/N_{\rm dof}$ are about 0.9 for shear only and joint fitting. Therefore, the data can be well fitted, and the best fitting curve curves within 1-$\sigma$ of the majority of data points.

In the Fig. \ref{fig:fit}, we show the contour maps with 1$\sigma$ (68.3\%) and 2$\sigma$ (95.5\%) C.L. of the free parameters, i.e. $\Omega_m$, $h$, $n_s$, $\sigma_8$, $c_1$, $c_2$ and four multiplicative calibration parameters $m$. The colored solid curves are and the black dashed lines are the the 1-D PDFs and fiducial values of the parameters, respectively. The red contours denote the results of the galaxy shear survey, and the blue contours are the joint constraint results. 

We can find that, all the best-fit values of free parameters are very close to the input values within 1-$\sigma$. For the joint survey, $\Omega_{\rm m} = 0.298^{+0.010}_{-0.011}$, $h = 0.702^{+0.020}_{-0.022}$, $n_{\rm s} = 0.947^{+0.020}_{-0.020}$ and $\sigma_8 = 0.802^{+0.013}_{-0.013}$ ( $\Omega_{\rm m} = 0.301^{+0.009}_{-0.009}$, $h = 0.698^{+0.018}_{-0.019}$, $n_{\rm s} = 0.950^{+0.013}_{-0.013}$ and $\sigma_8 = 0.898^{+0.009}_{-0.009}$ for $\ell_{\rm max}=3000$ ) in $1\sigma$ confidence level for the shear-shear power spectra with $\ell_{\rm max}=1000$, this constraints on the cosmological parameters are improved by 4\% to 9\% compared to only the galaxy shear survey. We find that joint fitting not only can shrink the probability contours but also improve the fitting biases. From Table \ref{tab:result}, one can see that after adding magnification, we can eliminate more than 50\% of the bias, especially when using the $\ell_{\rm max}=1000$ cut. 
In Figure \ref{fig:fit}, one can see that the magnification data mainly improve the joint constraint contours of IA model parameter ($c_1$/$c_2$) and cosmological parameter ($\sigma_8$/$\Omega_m$). 
Similar constraints can be achieved through galaxy clustering survey \citep{Ma23}.

\section{Summary and conclusion}
\label{conclusions}

In this work, we predict the LSST SNIa magnification measurement and CSST galaxy shear observation, and study the constraints of these observations on the cosmological parameters and intrinsic alignment model parameters. We divide the photo-$z$ distribution into five tomographic bins for LSST and four tomographic bins for CSST. For the shear measurement, in addition to noise, the intrinsic alignments are considered through the analysis of both tidal alignment term and tidal torquing term. Then, we calculate the auto- and cross-spectra of different tomographic redshift bins using the theoretical models, and generate correlated fiducial cosmic magnification maps and cosmic shear maps. 
We simulate the mock SNIa catalog and generate the mock magnification imaging data by analyzing the luminosity distance of SNIa in the tomographic redshift bins. For the cosmic shear, the effect of systematics and measure errors are included when generating the mock imaging data.

We estimate the auto- and cross-spectra of the mock SNIa magnification maps and mock galaxy shear maps by spherical harmonics transforming. However, the raw power spectra are contaminated by the mask and instrument noise. We set the values of the masked areas as zero, this treatment can introduce additional pseudo signal in the estimation of the power spectrum. Therefore, we use the mode coupling matrix $M_{\ell \ell^{\prime}}$ to correct this effect. Because cross-correlation can minimize the contamination from instrumental noise, we use the difference between the auto- and cross-correlation as noise to correct the raw power spectra. Then, we estimate the uncertainties by calculating the covariance matrix, and compare the measured angular power spectra with the fiducial values.

After obtaining the corrected power spectra, we adopt the MCMC technique to fit the power spectra using the theoretical model, and constrain the cosmological and systematical parameters. We study and compare two cases: one involving galaxy shape measurements only and the other involving joint SNIa magnification measurements together with galaxy shape measurements. We find that, by using only cosmic shear data, almost all the best-fit values of free parameters are very close to the input values within 1-$\sigma$, except for alignment model parameters $c_1/c_2$ and cosmological parameter. By adding magnification data, we are able to eliminate the $1\sigma$ bias in these three parameters.

\section*{Acknowledgements}
Y.C. and B.H. thank Zhengyi Wang for helpful discussion. Y.C. acknowledge the support of NSFC-12203003, the Project funded by China Postdoctoral Science Foundation 2022M720481. B.H. acknowledges the science research grants from the China Manned Space Project with No. CMS-CSST-2021-A12. J.Y. acknowledges the support from NSFC-12203084, the China Postdoctoral Science Foundation 2021T140451, and the Shanghai Post-doctoral Excellence Program grant No. 2021419. H.Z. acknowledges support by the National Key R\&D Program of China No. 2022YFF0503400 and China Manned Space Program. Some of the results in this paper have been derived using the healpy and HEALPix packages. 

\bibliography{CSSTxLSST}{}
\bibliographystyle{aasjournal}
\end{document}